\documentclass[final,5p,times,twocolumn]{article}

\usepackage{amssymb}
\usepackage{amsmath}
\usepackage{url}\usepackage{natbib}
\usepackage{url}
\usepackage{graphicx}

\begin{document}



\title{Galaxy image simplification using Generative AI} 


\author{Sai Teja Erukude, Lior Shamir \\ Department of Computer Science, Kansas State University \\ 1701 Platt St, Manhattan, KS 66506, USA} 

\date{}
\maketitle

\begin{abstract}
Modern digital sky surveys have been acquiring images of billions of galaxies. While these images often provide sufficient details to analyze the shape of the galaxies, accurate analysis of such high volumes of images requires effective automation. Current solutions often rely on machine learning annotation of the galaxy images based on a set of pre-defined classes. Here we introduce a new approach to galaxy image analysis that is based on generative AI. The method simplifies the galaxy images and automatically converts them into a ``skeletonized" form. The simplified images allow accurate measurements of the galaxy shapes and analysis that is not limited to a certain pre-defined set of classes. We demonstrate the method by applying it to galaxy images acquired by the DESI Legacy Survey. The code and data are publicly available. The method was applied to 125,000 DESI Legacy Survey images, and the catalog of the simplified images is publicly available. 
\end{abstract}



\section{Introduction}
\label{introduction}

Digital sky surveys are currently some of the most productive research instruments in astronomy \citep{kron1995digital,margony1999sloan,djorgovski2001exploration}, and in fact have revolutionized astronomy research \citep{ivezic2012galactic,tyson2012future}. Digital sky surveys such as the Sloan Digital Sky Survey (SDSS) \citep{york2000sloan}, the Panoramic Survey Telescope and Rapid Response System (Pan-STARRS) \citep{kaiser2002pan}, the Dark Energy Survey (DES) \citep{dark2016dark}, the Hyper Suprime-Cam (HSC) \citep{aihara2018hyper}, and the Kilo-Degree Survey (KiDS) \citep{de2013kilo} scan the sky automatically and store the data they collect to generate extremely large astronomical databases. These mature sky surveys are joined by new powerful research instruments such as the Vera Rubin Observatory \citep{ivezic2019lsst}, Euclid \citep{mellier2024euclid}, and Roman \citep{spergel2015wide}.

But to fully utilize the discovery power of these powerful digital sky surveys, there is a need for algorithms that can analyze billions of astronomical objects and turn them into usable data products. Namely. a large number of methods for automatic analysis of galaxy images have been proposed, many of them are based on machine learning  \citep{de2004machine,kasivajhula2007morphological,shamir2009automatic,almeida2010automatic,dieleman2015rotation,cheng2020optimizing,cheng2021galaxy,graham2019galaxy,shamir2011ganalyzer,ibrahim2018galaxy,kuminski2014combining,goddard2020catalog,schutter2015galaxy,sandeep2021analyzing,reza2021galaxy,li2022automatic,selim2016galaxy,shamir2013automatic,eassa2022automated,cecotti2020rotation,shamir2021automatic,ghadekar2022galaxy,shamir2014automatic,shamir2012automatic,semenov2025galaxy,yeganehmehr2025classification,abd2013intelligent,shamir2023outlier,kuminski2018hybrid,shamir2016morphology,bastanfard2019automatic,kwik2022galactic,huertas2021galaxy,ma2023galaxy,cheng2023lessons,urechiatu2024improved,cao2024galaxy,misra2020convoluted,walmsley2021galaxy,walmsley2023zoobot}. Some of these methods were applied to generate large catalogs of annotated galaxy images \citep{kuminski2016computer,dominguez2018improving,cheng2021galaxy,cheng2020optimizing,goddard2020catalog,walmsley2022galaxy,aussel2024euclid,bom2024extended}.

 One of the limitations of these algorithms is that they are normally based on supervised machine learning, such that they are trained and tested with a pre-defined set of annotated galaxies. These galaxies are labeled using a predefined set of classes. Then, the classifier can automate the classification of the galaxies into one of several distinct classes. For instance, such algorithms can classify galaxies by their broad morphology into elliptical and spiral galaxies, or into a stage in the Hubble sequence. Other tasks include the identification of gravitational lenses, peculiar galaxies, ring galaxies, and more.

Conceptualizing galaxy image analysis as an image classification problem allows to easily use the existing machine learning foundations and apply it to galaxy image analysis. That provides an accessible solution that enables the annotation of a large number of galaxies. Its application to the field of astronomy is therefore an effective use of the state-of-the-art computer science methodology.

However, the approach of borrowing a practice from the domain of object recognition also has limitations. Associating a galaxy with one of a set of pre-defined classes can be considered an oversimplification of galaxy morphology. For instance, galaxies within the same Hubble type can be different from each other, and therefore assigning them the same Hubble type does not fully reflect the differences between them. Galaxies may have different shapes for each arm, and that morphology is just partially reflected when annotating a galaxy into a pre-defined class. Also, classification of galaxies into predefined classes does not provide a precise quantitative measurements of the galaxy's shape elements. 

Also, conceptualizing galaxy morphology as a machine learning classification problem requires to predefine the classes and to pre-train a machine learning model. While such models can interpolate between the classes, their ability to analyze questions that are outside the scope of the training data is limited. Therefore, the research questions that can be studied largely depend on the design of the model and the training data.

Another limitation of using machine learning to classify galaxies is that these algorithms are based on complex data-driven rules that are very difficult to conceptualize. These algorithms are therefore tested empirically. But because convolutional neural networks use all pieces of information to separate between galaxies, they can be slightly biased by the data collection mechanisms \citep{ball2023ai, erukude2024identifying, erukude2024thesis}. Such biases are often difficult to notice \citep{ball2023ai}, and have also been identified in the application of convolutional neural networks to galaxy classification \citep{dhar2022systematic}.

Here we describe a new approach to galaxy morphology analysis. The approach makes use of generative AI to simplify a galaxy image, and transform it into its ``skeletonized" form that includes just the primitive shape of the galaxy arms. That analysis allows for the precision analysis of the galaxy shapes that can be adjusted to different scientific questions. 

\section{Data}
\label{data}

The data used in this study were taken from the 9th data release (DR9) of the Dark Energy Spectroscopic Instrument (DESI) Legacy Imaging Survey \citep{dey2019overview}. The DESI Legacy Survey covers $1.4 \cdot 10^4$ degrees$^2$ of the sky. The initial dataset used in this study included $1.3 \cdot 10^6$ galaxies taken randomly from the DESI Legacy Imaging Survey, such that the galaxies had {\it g} magnitude lower than 21. The initial images were downloaded from the DESI Legacy Survey server in the JPG format. The size of each image is 256$\times$256 pixels. To exclude point sources, only objects that were marked as exponential galaxies (‘REX’), exponential discs (‘EXP’), or de Vaucouleurs r1/4 profiles (‘DEV’) in DESI Legacy Survey DR9 were used.

\subsection{Preprocessing of DESI Legacy Survey images}
\label{preprocessing}

Naturally, the initial dataset included spiral galaxies, but also non-spiral galaxies and noisy images that the DESI Legacy Survey pipeline mistakenly labelled as galaxies. That required a first step of cleaning the data and selecting just the objects that are suitable for analysis.

\begin{figure}[ht]
    \centering
    \includegraphics[width=3.2in]{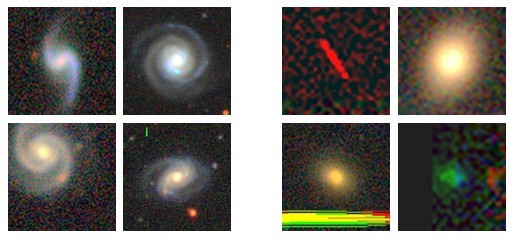}
    \caption{A sample of objects that are late-type (left) and objects that are not late-type galaxies (right) taken from DESI Legacy Survey images to train the ResNet50 model.}
    \label{fig_spiral_nonspiral}
\end{figure}

For that purpose, a new dataset was created from the DESI images to train a classifier that can distinguish between late-type galaxies and early-type galaxies, as well as noisy images. The training set consists of two classes: objects that are late-type galaxies, and objects that are not late-type galaxies. The training set contained 1,234 images, 617 in each class, while the validation set comprised 254 images, with 127 per class. Figure~\ref{fig_spiral_nonspiral} displays example images of the two classes of this dataset. The training, validation, and test datasets are made public and are available at \url{https://doi.org/10.6084/m9.figshare.28889549}.

To classify between the galaxies automatically, we employed a standard ResNet50 architecture \citep{he2016deep}, which is a robust and widely used Convolutional Neural Network. The neural network was pre-trained on ImageNet and fine-tuned using the training set of galaxies described above.

For improved fine-tuning, the Adam optimizer \citep{kingma2014adam} was used with a reduced learning rate of 0.0001. To retain the benefits of pretraining, only the final 20\% of the layers were left trainable, while the rest were frozen. We included callback functions, EarlyStopping, ModelCheckpoint, and ReduceLROnPlateau to monitor the learning process and save the best-performing model. 

\begin{figure}[ht]
    \centering
    \includegraphics[width=3in]{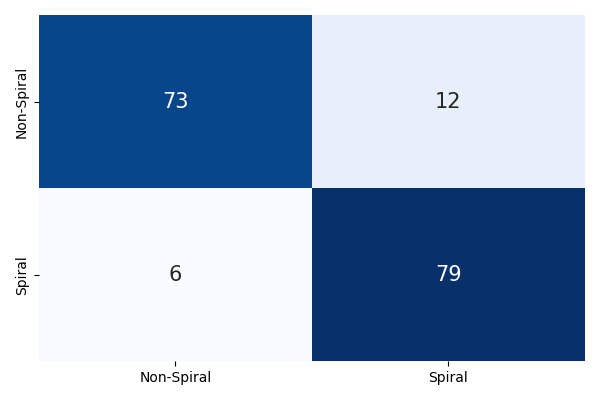}
    \caption{Confusion matrix of the ResNet50 model for classification between images of spiral galaxies and images of other objects that are not spiral galaxies.}
    \label{fig_resnet_confusion_matrix}
\end{figure}

\begin{figure}[ht]
    \centering
    \includegraphics[width=3.2in]{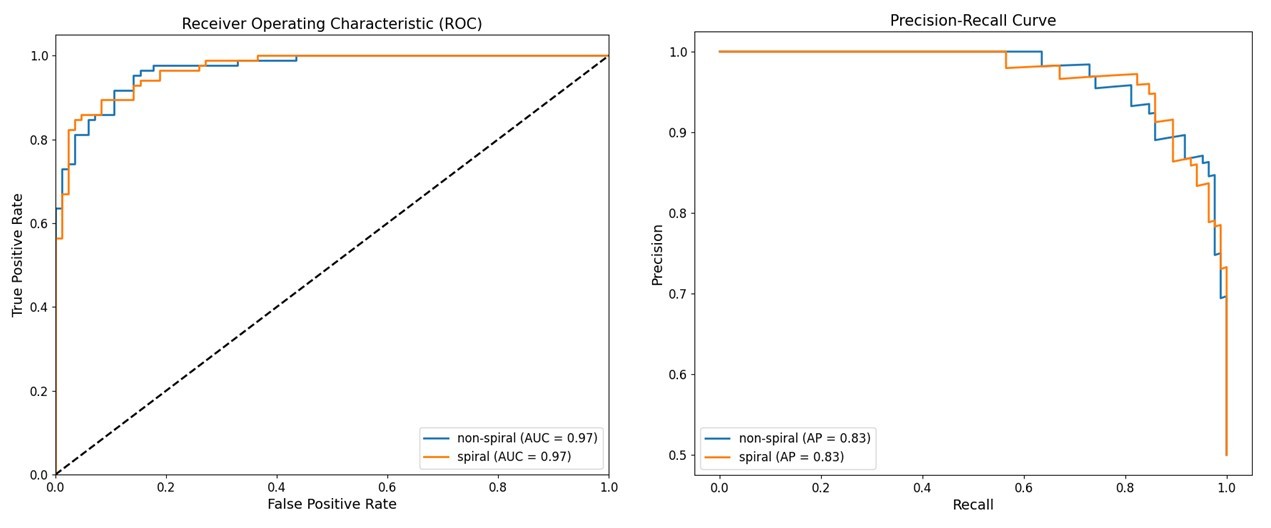}
    \caption{Evaluation metrics of the ResNet50 model: ROC curve, and the Precision-Recall curve.}
    \label{fig_resnet_metrics}
\end{figure}

The model was then evaluated on a hand-picked test dataset containing 170 samples, with 85 images from each class. It achieved an overall accuracy of $\approx$ 89.5\% and an area under the curve (AUC) of around $\approx$ 0.97. Figure~\ref{fig_resnet_confusion_matrix} shows the confusion matrix, and Figure~\ref{fig_resnet_metrics} shows the ROC curve and the Precision-Recall curve. Applying the neural network led to a dataset of 125,000 objects identified as late-type galaxies.

\subsection{Data to train the cGAN}
\label{gan_training_data}

\begin{figure}[ht]
    \centering
    \includegraphics[width=3.2in]{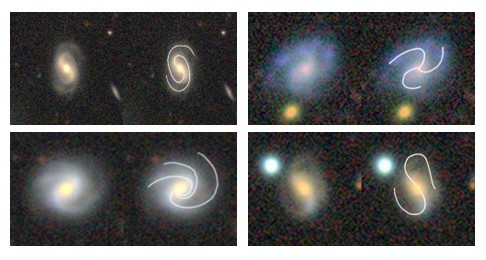}
    \caption{Pairs of training images used to train the cGAN. The left is the original image, and the right is the target image that the cGAN is trained to generate.}
    \label{cgan1_training_samples}
\end{figure}

The network was trained on a manually annotated dataset of spiral galaxy images, such that each galaxy image in the dataset was paired with a manually annotated image. The training data include pairs of images, such that one image is the original galaxy image, and the other image is an image annotated manually with white lines on the arms of the galaxy. These pairs of images are used to train the cGAN to generate a simple ``skeletonized" image from the original galaxy images. The annotations of the training set were done manually with a Samsung Tablet and its pen. Figure~\ref{cgan1_training_samples} shows several examples of the training data.

Data augmentation techniques such as blurring, zooming, and flipping, combined with rotation and brightness adjustment, were applied to increase the size and robustness of the dataset.  Blurring was applied using a Gaussian blur with a kernel size of 5$\times$5, and a standard deviation of 0 was used to reduce noise and smooth fine details in the images. To introduce spatial variation, each image was flipped horizontally and then rotated by 90$^o$ clockwise. A zoom effect was simulated by performing a center crop at 75\% of the image, and then resizing it back to the original size using bilinear interpolation. Additionally, brightness was increased using OpenCV’s convertScaleAbs() function with an alpha value of 1.2 (adjusting contrast) and a beta value of 10 (increasing brightness). The code used in the pre-processing pipeline is available in \url{https://github.com/SaiTeja-Erukude/galaxy-image-simplification-using-genai}.

After applying the data augmentation, a set of 4,180 pairs was used for the training of the cGAN. The training images are available at \url{https://doi.org/10.6084/m9.figshare.28889549}.

\section{A cGAN-based method for simplification of galaxy images}
\label{method}

A Generative Adversarial Network \citep{goodfellow2014generative} is a deep learning architecture used to generate synthetic data that mimics real-world data. It consists of two competing neural networks: a generator and a discriminator. As its name suggests, the generator takes input data and generates synthetic data samples that are optimized during training to be indistinguishable from real data. The discriminator, on the other hand, acts like a cop trying to catch the forger. It tries to predict whether the generated data is real or fake. Both networks are then trained in an adversarial manner. The generator learns from the discriminator's feedback to produce more realistic outputs, and the discriminator learns from its mistakes, improving its ability to distinguish between real and fake data. This process is done iteratively, with the generator and discriminator both getting better at their jobs.

We employed a Conditional Generative Adversarial Network (cGAN) to identify and delineate the arms of galaxies. cGAN is an extension of GAN, where the generator network incorporates additional input information or conditioning variables that influence the data generation process. 

\begin{figure*}[ht]
    \centering
    \includegraphics[width=5in]{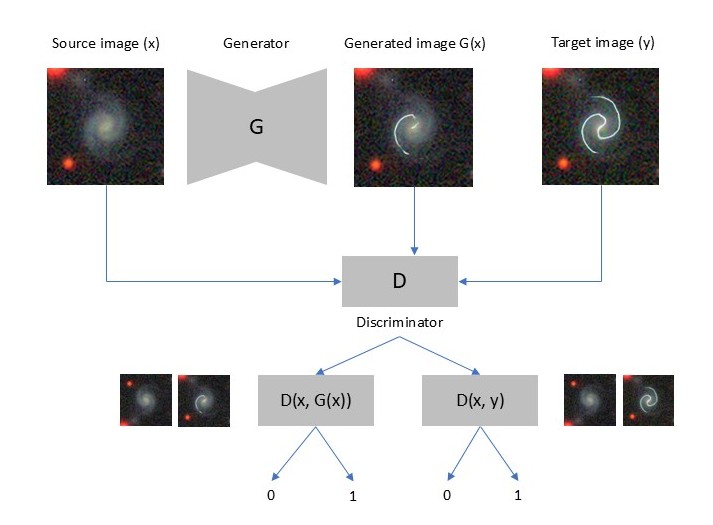}
    \caption{Schematic overview of the architecture of the applied cGAN.}
    \label{fig_cgan_working}
\end{figure*}

\cite{isola2018imagetoimagetranslationconditionaladversarial} explain how cGANs can be used as a general-purpose solution to image-to-image translation problems. Here we follow a similar approach, where the networks are trained on the input and output images. This architecture allows the networks to learn and transform an input image from one style to another. In the case of galaxy image simplification, we convert an input galaxy image to the corresponding galaxy image with delineated arms, as shown in Figure~\ref{fig_cgan_working}.

\subsection{cGAN Architecture}

\begin{figure*}[ht]
    \centering
    \includegraphics[width=5in]{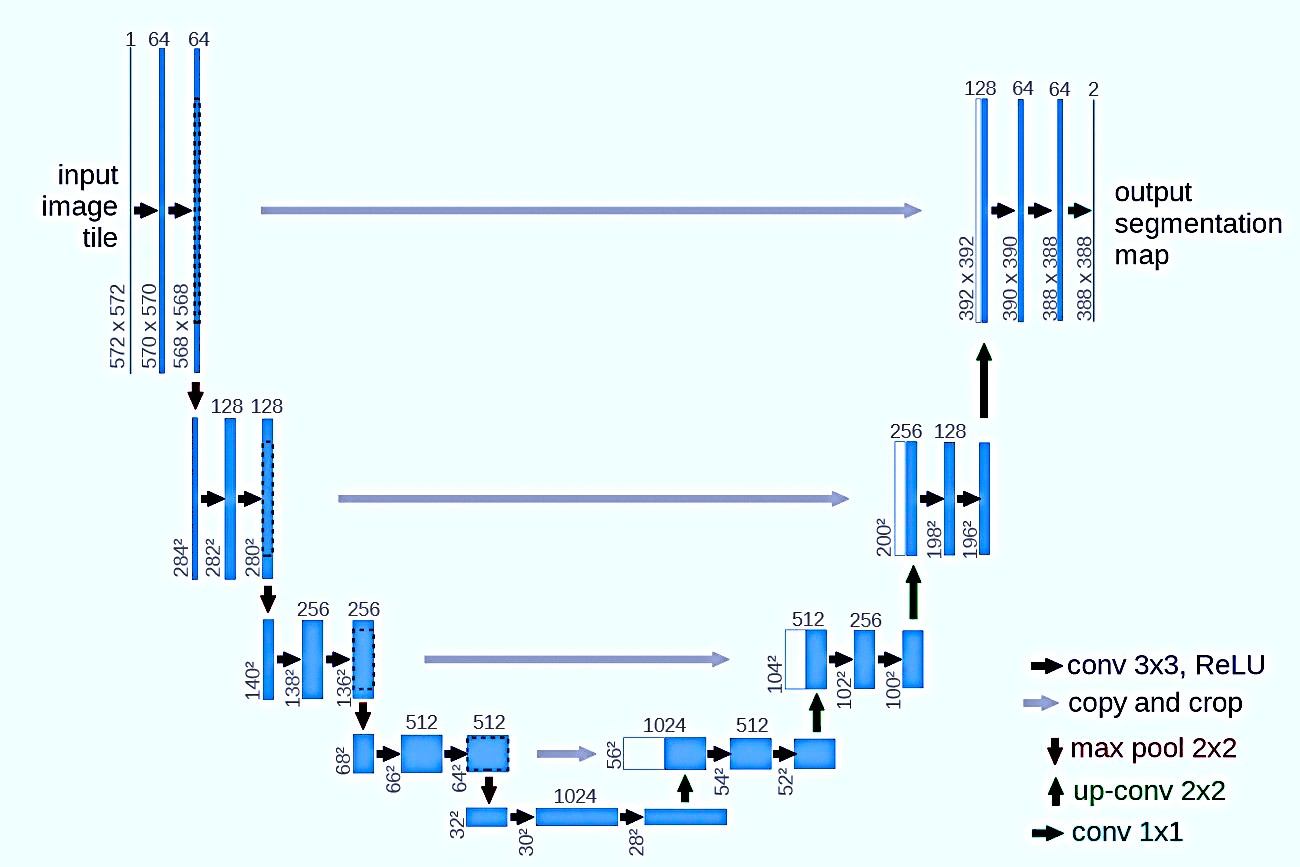}
    \caption{U-Net based generator architecture.}
    \label{fig_generator_architecture}
\end{figure*}

The generator follows a U-Net-based architecture \citep{ronneberger2015u}, which allows for the preservation of spatial details through skip connections. The network consists of a contracting path (encoder) and an expansive path (decoder) arranged in a distinctive U-shaped structure, where skip connections directly link corresponding layers between the encoder and decoder paths, depicted in Figure~\ref{fig_generator_architecture}. This architectural design enables the network to effectively combine high-level contextual information with low-level spatial details, facilitating the generation of high-fidelity outputs. The GAN was implemented using the Keras library.

\begin{figure*}[ht]
    \centering
    \includegraphics[width=5in]{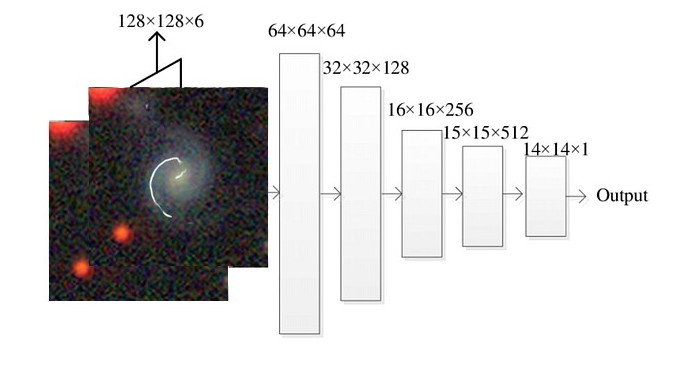}
    \caption{PatchGAN discriminator architecture.}
    \label{fig_discriminator_architecture}
\end{figure*}

The discriminator employs a PatchGAN architecture \citep{demir2018patch}, shown in Figure \ref{fig_discriminator_architecture}. The PatchGAN discriminator operates by classifying individual N$\times$N patches of the input image rather than the entire image at once. This design choice is particularly effective for capturing high-frequency details and textures, crucial for maintaining local image authenticity. The PatchGAN discriminator processes the image convolutionally, treating it as a Markov random field and assuming independence between pixels separated by more than a patch diameter. This approach helps in preserving fine-grained details and textures in the generated outputs. Because the output is expected to be in the form of thin lines, an architecture that is sensitive to the fine details is critical.

The loss $L_g$ is the weighted sum of the adversarial loss $L_{gan}$ (Binary Cross Entropy) and the $L_1$ loss (Mean Absolute Error), as shown in Equation~\ref{loss_function}. The Adversarial loss $L_{gan}$ is the discriminator loss, which gets lower when the generator produces outputs that ``fool" the discriminator. The Reconstruction loss ($L_1$) is determined by the mean absolute difference between the real images and the generated images. It gets lower when the generated image is more similar to the target image. $\lambda$ is the hyperparameter set to 0.5 that controls the strength of the $L_1$ regulator.

\begin{equation}
L_g = L_{gan} + \lambda \cdot L_1.
\label{loss_function}
\end{equation}

\subsection{Training the model}

The model was trained using the dataset of image pairs described in Section~\ref{gan_training_data}. Each training sample is a pair of images such that one image is the original galaxy image. The other image is the annotated image, which is the target image that the cGAN is trained to generate automatically from the source image. As explained in Section~\ref{data}, each target image is the source image, such that the galaxy arms are marked with white lines. We used the Adam optimizer \citep{kingma2014adam} with a learning rate of 0.0002, batch size of 1, and trained for 100 epochs. The output of the 50th epoch was used, as will be explained in Section~\ref{results}. 

\subsection{Post-processing of fixing ``broken" simplified arm lines}
\label{post_processing}

Although the cGAN-generated images were mostly of good quality, some had broken or shaky arm lines with a low degree of confidence. To address the problem, several image processing techniques were considered to connect and smooth arm lines. Structure from Motion techniques that were attempted, for instance, included edge detection, Bresenham's algorithm, and some morphological operations. However, these are mostly simple distance-dependent techniques (i.e., specifying a fixed distance threshold like `X' pixels) to connect gaps between a point to nearby line segments. 

\begin{figure}[ht]
    \centering
    \includegraphics[width=3.2in]{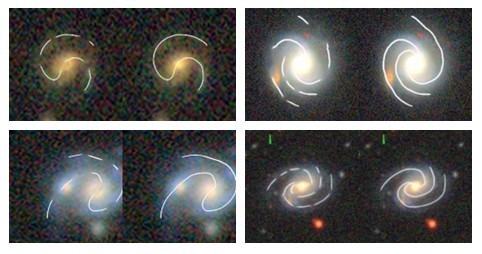}
    \caption{Pairs of training images used to train the post-processing cGAN. The left is the cGAN1-generated image with broken lines, and the right is the target image that the cGAN2 is trained to generate.}
    \label{fig_cgan2_training_samples}
\end{figure}

To connect the broken line in a manner that is sensitive to the content of the entire image, a second cGAN was used. That cGAN was trained using the same configuration and architecture as the first GAN, employing the Adam optimizer with a learning rate of 0.0002, a batch size of 1, and running for 100 epochs. As with the first model, the version from the 50th epoch was selected for its superior generalizability. Figure \ref{fig_cgan2_training_samples} presents a selection of training samples used to train this post-processing cGAN. A set of 3,172 pairs was used for the training of the cGAN, and the training images are available at \url{https://doi.org/10.6084/m9.figshare.28889549}. 

\begin{figure*}[ht]
    \centering
    \includegraphics[width=5in]{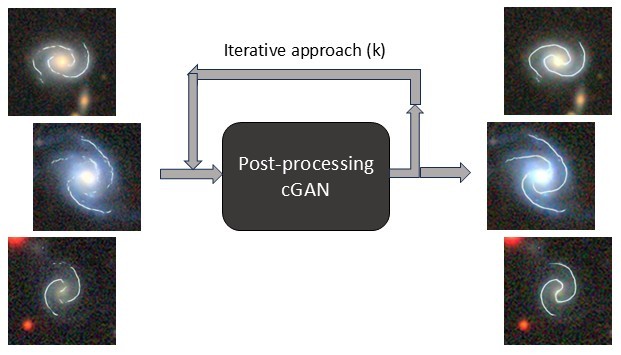}
    \caption{Post-processing cGAN with an iterative approach to join broken arm lines}
    \label{fig_cgan2_application}
\end{figure*}

As shown in Figure~\ref{fig_cgan2_application}, an iterative approach was adopted to address the broken arm lines. We experimented with different numbers of iterations, ranging from 1 through 10, and concluded that iterating the process five times yielded the most desirable results. Each pass incrementally improved the output and gradually joined the broken lines. The improvements become noticeable at the end of the 5th iteration.

\subsection{Separating the ``Skeletonized" arm lines}

\begin{figure}[ht]
    \centering
    \includegraphics[width=3.2in]{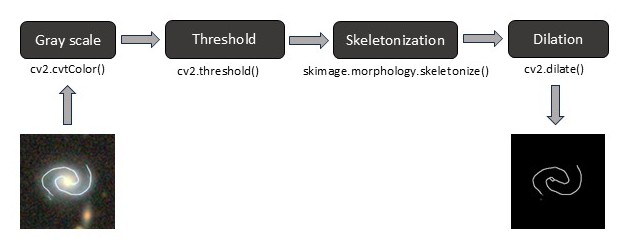}
    \caption{Image processing workflow to skeletonize the arm lines.}
    \label{fig_skeletonization_process}
\end{figure}

To accurately separate the arm lines from the processed images, we employed a systematic image processing workflow using OpenCV \citep{bradski2000opencv}, outlined in Figure~\ref{fig_skeletonization_process}. The process starts by converting the image to grayscale for easier processing. A simple binary threshold is then applied, where pixel values above a threshold of 200 are set to white (255), and those below it to black (0), effectively highlighting the white lines in the image. 

Skeletonization is then performed using skimage \citep{van2014scikit}, which reduces the white lines to their thinnest possible representation. To enhance the visibility of the skeletonized lines, dilation is employed with a kernel size of 2$\times$2. The final result is then stored, providing a clean and enhanced representation of the isolated arm lines. This sequence of operations provides a robust method for isolating and enhancing white lines in an image. This skeletonization workflow code is made public and available at \url{https://github.com/SaiTeja-Erukude/galaxy-image-simplification-using-genai/blob/main/postprocess/skeletonize.py}.

\section{Experimental results}
\label{results}

The ultimate goal of the method is to be able to generate catalogs of simplified galaxy images. Researchers can use the code developed for this project to train a cGAN based on the data of their choice. Then, the cGAN can be applied to the data to simplify a large number of galaxies. The method can be adjusted to different telescopes,  but it requires a training set of pairs of original and skeletonized images taken with that specific telescope.

The evaluation of GANs is known to be challenging \citep{shmelkov2018good,borji2022pros}, as GANs often generate complex data that is difficult to quantify in a manner that reflects the true ability of the GAN to generate data. For instance, ``deep fake" GANs are evaluated by their ability to deceive a person, which can be tested by cognitive tests rather than numerical analysis \citep{arora2021review,ben2024overview}. Another example is style GANs, which are also evaluated by cognition \citep{zhang2017style,lang2021explaining}. Simple evaluation such as image-to-image comparison between generated skeletonized images and ground truth images require the generated arm pixels to be in exact match with the ground truth arm pixels, making such comparison non-trivial. 

\begin{figure}[ht]
    \centering
    \includegraphics[width=3.2in]{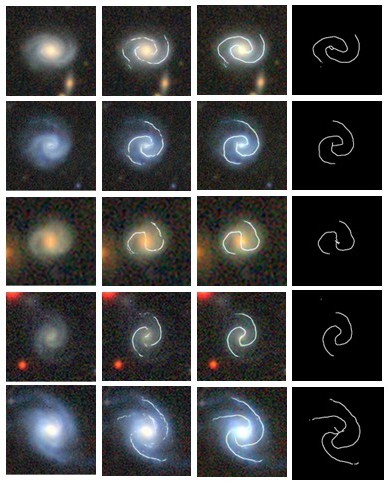}
    \caption{Example images as processed by the method. The original images are converted by the cGAN into images with the white lines highlighting the arms. A correction for broken lines is applied, followed by binary thresholding to create images that show just the shape of the arms.}
    \label{fig_results}
\end{figure}

Given that there is no existing benchmark or method for assessing the performance of galaxy simplification algorithms, the performance was evaluated by manual inspection of the original galaxy images compared to the generated galaxy images. Figure~\ref{fig_results} shows several examples of the original galaxies and the simplified galaxy images generated by the cGAN.

As the figure shows, the images at the end of the process are binary masks that show just the galaxy arms. Such images can be easily analyzed to study the shapes of the galaxy arms in a manner that is not necessarily taken into consideration when classifying the galaxy images into one of a pre-defined set of distinct classes. While the figure shows just a few examples, these examples are representative of the entire catalog. The catalog is available online and described in Section~\ref{catalog}.

\section{A catalog of 125,000 simplified galaxy images from DESI Legacy Survey}
\label{catalog}

\begin{figure}[ht]
    \centering
    \includegraphics[width=3.2in]{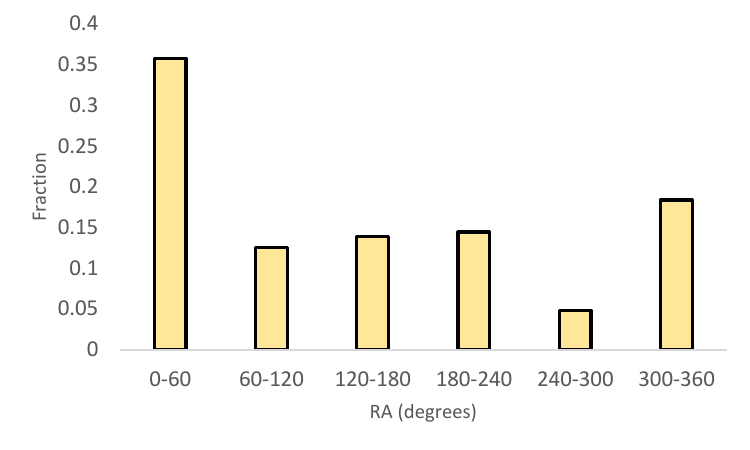}
    \caption{Distribution of the RA of the galaxies in the catalog.}
    \label{ra_distribution}
\end{figure}

The method was applied to $1.25\cdot 10^5$ galaxies from the DESI Legacy Survey to provide a catalog of simplified galaxy images. The galaxies are in the Southern hemisphere, imaged by the Dark Energy Camera (DECam) of the Blanco telescope in Cerro Tololo, Chile. The catalog is available for download at \url{https://doi.org/10.6084/m9.figshare.28889549}. For disk space and download speed considerations, the galaxy images in the catalog are in the JPG format. Each image is in the dimensionality of 256$\times$256 pixels. Figure~\ref{ra_distribution} shows the RA distribution of the galaxies.

The galaxies in the catalog can be identified by the file names. Each file name contains four numbers, separated by an underscore character. These numbers are the brick ID, the object ID, the right ascension, and the declination. The object can be identified by the combination of the brick ID and object ID, or by the combination of the right ascension and declination. The brick ID and object ID correspond to data release (DR) 8 of the DESI Legacy Survey.

For instance, a filename of a processed galaxy image in the catalog can be: ``45056\_2404\_6.372944834749819\_-59.76292961103172.jpg". In this case, the galaxy is object 2404 in brick 45056, and is at coordinates $(\alpha=6.372944834749819^o,\delta=-59.76292961103172^o)$.

\begin{figure}[ht]
    \centering
    \includegraphics[width=3.2in]{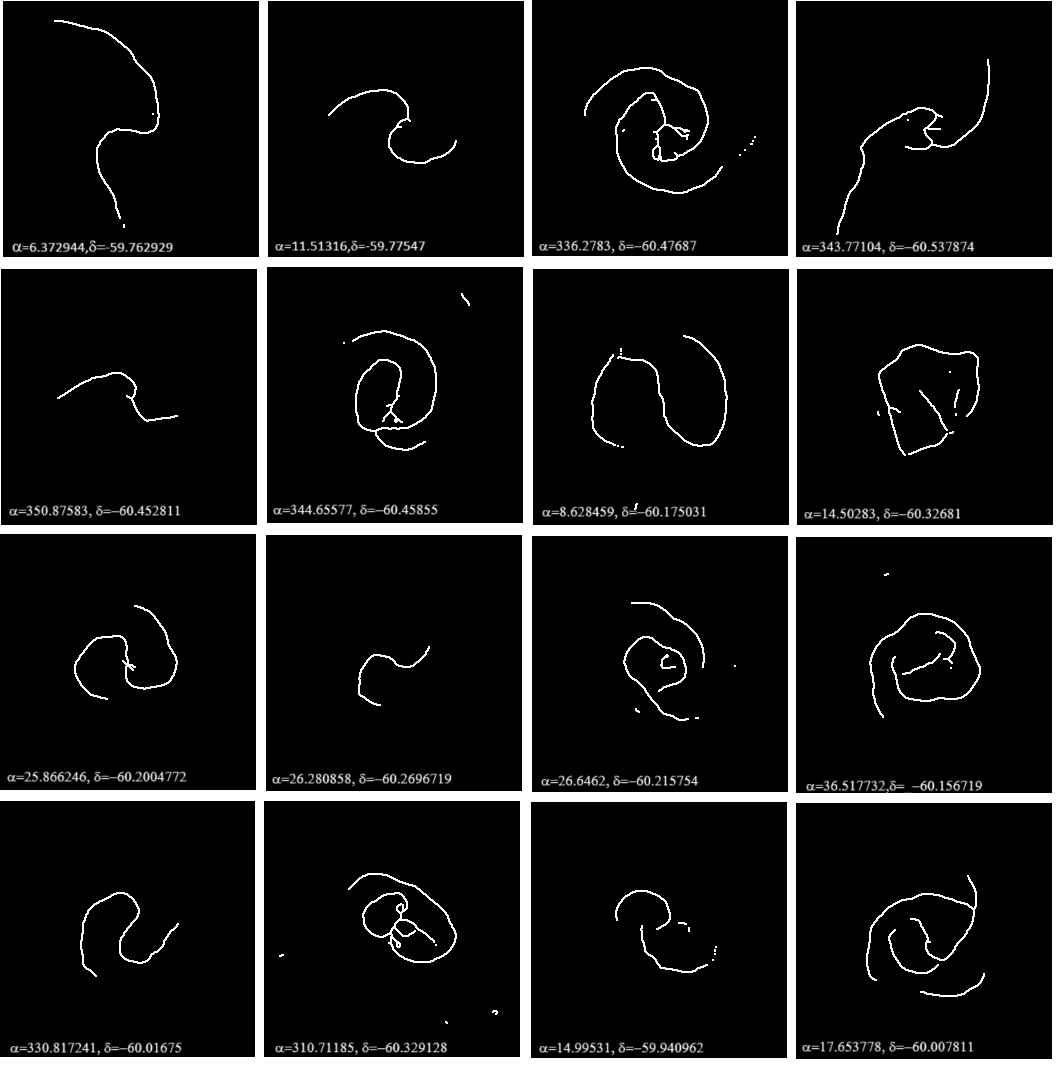}
    \caption{Example images from the catalog of simplified DESI Legacy Survey images.}
    \label{catalog_example}
\end{figure}

\begin{figure}[ht]
    \centering
    \includegraphics[width=3.2in]{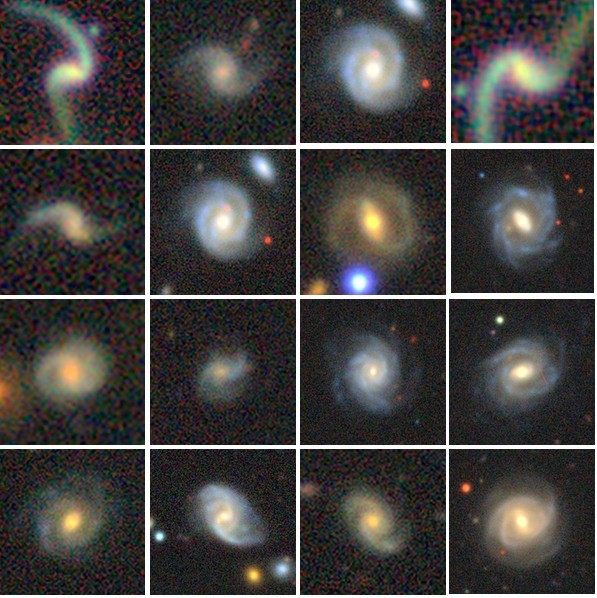}
    \caption{Example source images from DESI Legacy Survey images.}
    \label{catalog_example_sources}
\end{figure}

Figure~\ref{catalog_example} shows examples of random galaxies taken from the catalog. The coordinates of each galaxy are taken from the file names. These images are far easier to process to analyze different aspects of the galaxy arms. Figure~\ref{catalog_example_sources} shows the source images of the galaxies.

\section{Conclusion}
\label{conclusion}

The availability of robotic telescopes enables digital sky surveys with high-throughput pipelines of astronomical images. Among other data, these digital sky surveys collect and store billions of galaxy images. While these images can enable paramount scientific discoveries about the past and future Universe, these data are not always fully utilized. One of the reasons for the under-utilization of these data is the complexity of the images, making it far more difficult to query and explore compared to data provided in numerical forms, such as photometry or spectroscopy data.

Among all astronomical objects, late-type galaxies are the objects with the most complex morphology and are therefore the most difficult to fully analyze. Existing solutions include the utilization of machine learning to classify these galaxies into sets of pre-defined classes. 

That approach can provide an effective, broad analysis of the galaxies, but it depends on pre-defined classes determined by those who prepare the catalog. Researchers who are interested in other morphological features that were not considered by those who prepared the catalog will not be able to use the analyzed data. Moreover, a broad separation of galaxies into morphological classes will not allow specific measurements related to the accurate shapes of the arms.

By using generative artificial intelligence, we introduce an approach of converting a complex galaxy image into its ``skeletonized" form. After applying the first step of simplification of the galaxy images, the simplified form allows to use a secondary analysis based on the requirements of the researcher. Because the galaxies are in a simple form of binary images, the analysis of the arm can be performed easily by simple image analysis.

Possible application of the method can include characteristics of the shape of the arms such as the degree of the curve, separating between straight arms or highly curved arms. The length of the arms can also be measured and used to identify the difference between the shortest and longest arms, the average length of the arms, and more. The number of arms of each galaxy is also a characteristic that is difficult to determine when using the original galaxy image. The number of spiral arms has been shown to be linked to star
formation rate and stellar mass \citep{smith2022galaxy}.

Because the shapes of the galaxy arms are linked to the distribution of the mass of the galaxy, the method described here can allow more accurate analysis of the mass distribution within the galaxy. When the redshift of the galaxy is known, the method can also be used to analyze galaxy evolution by profiling the changes in the galaxy arms over time. When analyzing a large number of galaxies within a certain sector in the sky, such analysis can be used to analyze the structure and distribution of galaxies within cluster, filaments, or the large-scale structure of the Universe.

The simplification of galaxy images can also be used for improving the performance of algorithms that annotate galaxy images. The annotation of the simplified images might make the classification of galaxies easier, and therefore can improve the accuracy of these methods. The combination of skeletonizing the galaxy images with their annotation can be studied in the future.

The downside of the method is that the conversion of the galaxy images into the simplified form loses substantial information. That includes the bulge of the galaxy or the thickness of the arms, which can be important for studying the galaxies. Satellite galaxies and star clusters might also be lost in the process, leaving just the shapes of the arms. Another disadvantage is that like many other machine learning methods, the GAN requires annotated images for training. The annotations of the images might be subjective by the perception of the person preparing the training set, which can consequently affect the analysis of the GAN.

The advantage of the method is that it maintains the shapes of the galaxy arms and other basic features of the galaxies, providing a new way of studying galaxy morphology. Therefore, the method is useful for researchers who wish to study questions related to the shape of the galaxy arms. It might not be relevant to other features of the galaxy, or to early-type galaxies that do not have arms.

The method was applied to the DESI Legacy Survey to provide a catalog of 1.25$\cdot 10^5$ simplified galaxy images. Given the continuous advancement of digital sky surveys and their increasing importance in modern astronomy, it is expected that such algorithms will become pivotal in the process of turning data collected by digital sky surveys into knowledge and scientific discoveries.

\section*{Data availability}

The catalog of simplified galaxy images as well as annotated training data for the cGAN can be downloaded at \url{https://doi.org/10.6084/m9.figshare.28889549}. Code used in this project is available at \url{https://github.com/SaiTeja-Erukude/galaxy-image-simplification-using-genai}. 

\section*{Acknowledgments}

We would like to thank the knowledgeable anonymous reviewer for the insightful comments. The research was supported in part by NSF grant number 2148878.

The Legacy Surveys consist of three individual and complementary projects: the Dark Energy Camera Legacy Survey (DECaLS; Proposal ID 2014B-0404; PIs: David Schlegel and Arjun Dey), the Beijing-Arizona Sky Survey (BASS; NOAO Prop. ID 2015A-0801; PIs: Zhou Xu and Xiaohui Fan), and the Mayall z-band Legacy Survey (MzLS; Prop. ID 2016A-0453; PI: Arjun Dey). DECaLS, BASS and MzLS together include data obtained, respectively, at the Blanco telescope, Cerro Tololo Inter-American Observatory, NSF’s NOIRLab; the Bok telescope, Steward Observatory, University of Arizona; and the Mayall telescope, Kitt Peak National Observatory, NOIRLab. Pipeline processing and analyses of the data were supported by NOIRLab and the Lawrence Berkeley National Laboratory (LBNL). The Legacy Surveys project is honored to be permitted to conduct astronomical research on Iolkam Du’ag (Kitt Peak), a mountain with particular significance to the Tohono O’odham Nation.

NOIRLab is operated by the Association of Universities for Research in Astronomy (AURA) under a cooperative agreement with the National Science Foundation. LBNL is managed by the Regents of the University of California under contract to the U.S. Department of Energy.

This project used data obtained with the Dark Energy Camera (DECam), which was constructed by the Dark Energy Survey (DES) collaboration. Funding for the DES Projects has been provided by the U.S. Department of Energy, the U.S. National Science Foundation, the Ministry of Science and Education of Spain, the Science and Technology Facilities Council of the United Kingdom, the Higher Education Funding Council for England, the National Center for Supercomputing Applications at the University of Illinois at Urbana-Champaign, the Kavli Institute of Cosmological Physics at the University of Chicago, Center for Cosmology and Astro-Particle Physics at the Ohio State University, the Mitchell Institute for Fundamental Physics and Astronomy at Texas A\&M University, Financiadora de Estudos e Projetos, Fundacao Carlos Chagas Filho de Amparo, Financiadora de Estudos e Projetos, Fundacao Carlos Chagas Filho de Amparo a Pesquisa do Estado do Rio de Janeiro, Conselho Nacional de Desenvolvimento Cientifico e Tecnologico and the Ministerio da Ciencia, Tecnologia e Inovacao, the Deutsche Forschungsgemeinschaft and the Collaborating Institutions in the Dark Energy Survey. The Collaborating Institutions are Argonne National Laboratory, the University of California at Santa Cruz, the University of Cambridge, Centro de Investigaciones Energeticas, Medioambientales y Tecnologicas-Madrid, the University of Chicago, University College London, the DES-Brazil Consortium, the University of Edinburgh, the Eidgenossische Technische Hochschule (ETH) Zurich, Fermi National Accelerator Laboratory, the University of Illinois at Urbana-Champaign, the Institut de Ciencies de l’Espai (IEEC/CSIC), the Institut de Fisica d’Altes Energies, Lawrence Berkeley National Laboratory, the Ludwig Maximilians Universitat Munchen and the associated Excellence Cluster Universe, the University of Michigan, NSF’s NOIRLab, the University of Nottingham, the Ohio State University, the University of Pennsylvania, the University of Portsmouth, SLAC National Accelerator Laboratory, Stanford University, the University of Sussex, and Texas A\&M University.

BASS is a key project of the Telescope Access Program (TAP), which has been funded by the National Astronomical Observatories of China, the Chinese Academy of Sciences (the Strategic Priority Research Program ''The Emergence of Cosmological Structures” Grant  XDB09000000), and the Special Fund for Astronomy from the Ministry of Finance. The BASS is also supported by the External Cooperation Program of Chinese Academy of Sciences (Grant 114A11KYSB20160057), and Chinese National Natural Science Foundation (Grant 12120101003,  11433005).

The Legacy Survey team makes use of data products from the Near-Earth Object Wide-field Infrared Survey Explorer (NEOWISE), which is a project of the Jet Propulsion Laboratory/California Institute of Technology. NEOWISE is funded by the National Aeronautics and Space Administration.

The Legacy Surveys imaging of the DESI footprint is supported by the Director, Office of Science, Office of High Energy Physics of the U.S. Department of Energy under Contract No. DE-AC02-05CH1123, by the National Energy Research Scientific Computing Center, a DOE Office of Science User Facility under the same contract; and by the U.S. National Science Foundation, Division of Astronomical Sciences under Contract No. AST-0950945 to NOAO.

\bibliographystyle{apalike} 
\bibliography{galaxy_gan}

\end{document}